\documentstyle[aps,epsf]{revtex}
\def\lsim{\mathrel{\rlap{
\lower4pt\hbox{\hskip-3pt$\sim$}}
    \raise1pt\hbox{$<$}}}     
\def\gsim{\mathrel{\rlap{
\lower4pt\hbox{\hskip-3pt$\sim$}}
    \raise1pt\hbox{$>$}}}     

\begin{document}
\input{epsf}

\draft
\twocolumn[\columnwidth\textwidth\csname@twocolumnfalse\endcsname
\title{ Directed Flow of Baryons  in Heavy-Ion Collisions }
\author { Yu.B.~Ivanov$^{ a,b}$, 
E.G.~Nikonov$^{a,c }$,  
W.~N\"orenberg$^{a}$, \\
A.A.~Shanenko$^{c}$, 
 and   V.D.~Toneev$^{a,c} $ } 
\address{ $^{a} $
 Gesellschaft f\"ur Schwerionenforschung mbH, Planckstr.$\!$ 1,
64291 Darmstadt, Germany}
\address{ $^{b} $
Kurchatov Institute, Kurchatov sq.$\!$ 1, Moscow 123182, Russia}
\address{ $^{c}$  Joint Institute for Nuclear Research,
  141980 Dubna, Moscow Region, Russia } 

\maketitle

\begin{abstract}
The collective motion of nucleons from high-energy heavy-ion collisions 
is analyzed within a relativistic two-fluid model for different
equations of state (EoS). As function of beam energy the theoretical
 slope parameter $F_y$ of the differential directed flow is in  
good agreement with experimental data, when calculated 
for the QCD-consistent EoS described by the statistical mixed-phase
model. Within this model, which takes the deconfinement phase
transition into account,  the excitation function of the directed flow 
$\left< P_x\right> $  turns out to be a smooth function in 
the whole range from SIS till SPS energies. 
This function is close to that for pure hadronic EoS and
exhibits no minimum predicted earlier for a two-phase bag-model EoS.
Attention is also called to a possible formation of nucleon antiflow
($F_y<0$) at energies $\gsim 100$ A$\cdot$GeV.
\end{abstract}


\addvspace{5mm}]

\narrowtext

\section{Introduction}
Collective flows of various types (radial, directed, elliptic,...)
observed experimentally in heavy-ion collisions reveal a space--momentum
correlated motion of strongly interacting nuclear matter. This collective
motion is essentially caused by the pressure gradients arising
during the time evolution in the collision, and hence opens a 
promising way for obtaining information  on the equation of state (EoS) 
and, in particular, on a possible phase transition.
Recently, this feature has stimulated  a large number
 of experimental and theoretical investigations on
 flow effects (cf. review articles \cite{BMS98,Dan99}).

Manifestations of the deconfinement
phase transition have been considered already some time ago
by Shuryak, Zhirov~\cite{SZ79} and van Hove~\cite{V83}. 
Since a phase transition slows down
 the time evolution of the system due to {\em softening} of the EoS,
the authors expect around some critical incident energy a 
remarkable loss  
 of correlation between the observed particle momenta and the
 reaction plane, and hence a reduction of the directed flow. 
 Assuming a first-order phase transition 
  Hung, Shuryak~\cite{HS95} and  Rischke et al.~\cite{R96}
   have recently obtained   quantitative 
predictions for heavy-ion collisions.  
For an expanding fireball   Hung and Shuryak
 expect the  {\em softest point} effect
 around $E_{lab}=30$ A$\cdot$GeV. Within a one-fluid hydrodynamic 
model  Rischke et al.~show that the excitation 
function of the directed flow exhibits a deep minimum near 
$E_{lab}= 6$ A$\cdot$GeV. However, preliminary experimental 
results~\cite{E895} in this energy range do  not  confirm these 
predictions. In the following, we report on a study of  the
 directed flow within a two-fluid hydrodynamic model~\cite{MRS91} 
for  the statistical mixed-phase EoS~\cite{NST98,TNS98} which is
adjusted to available lattice QCD data.

\section{Equation of state within the mixed-phase model}

 Our consideration is essentially based on the
 recently proposed Mixed-Phase (MP) model~\cite{NST98,TNS98}
 which is consistent with 
available QCD lattice data \cite{karsch}.
The underlying  assumption of the MP model
is that  unbound quarks and gluons {\it may coexist}
with hadrons forming a {\it homogeneous}
quark/gluon--hadron phase. Since the mean
distance between hadrons and quarks/gluons in this mixed phase
may be of the same order as that between hadrons, the interaction
between all these constituents
 (unbound quarks/gluons and hadrons) plays an important 
role and defines the order of the phase transition.

Within the MP model \cite{NST98,TNS98} the
effective Hamiltonian is expressed in the quasiparticle
approximation with density-dependent mean-field interactions.
 Under quite general requirements of  confinement
for color charges,
the mean-field potential of  quarks and gluons
is approximated by
\begin{equation}
U_q(\rho)=U_g(\rho)={A\over\rho^{\gamma}}~;
 \ \ \ \gamma >0
\label{eq6}
   \end{equation}
with  {\it the total density of quarks and gluons}
$$
\rho=\rho_q + \rho_g +\sum\limits_{j}\;\nu_j\rho_{j}~,
$$
where $\rho_q$ and  $\rho_g$ are
the densities of unbound quarks and gluons outside of hadrons,
while $\rho_{j}$ is the density of hadron type $j$ and $\nu_j$ is the
number of valence quarks inside.
The presence of the total density $\rho$ in (\ref{eq6})
implies interactions
between all components of the mixed phase.
The  approximation (\ref{eq6})  mirrors two important
limits of the QCD interaction. For 
 $\rho \to 0$, the interaction potential approaches infinity,
{\em i.e.}  an infinite energy is necessary to create an isolated
quark or gluon, which simulates the confinement
of color objects. In the other extreme case of large energy density
corresponding to $\rho \to \infty$, we have  $U_q=U_g=0$ which is 
consistent with asymptotic  freedom.

The use of the density-dependent  potential (\ref{eq6}) for quarks
 and the hadronic potential, described by a
modified non-linear mean-field model~\cite{Zim},  requires certain
constraints to be fulfilled, which are related to 
thermodynamic consistency~\cite{NST98,TNS98}. For the
chosen form of the Hamiltonian these conditions require
that  $U_g(\rho)$ and $U_q(\rho)$ do not depend on temperature. 
From these conditions one also obtains a  
form for the quark--hadron potential~\cite{NST98}.

A detailed study of the pure gluonic $SU(3)$ case with a 
first-order phase transition allows one to fix the values of the
parameters as $\gamma =0.62$ and $\displaystyle
A^{1/(3\gamma+1)} = 250$ MeV.
These values are then used for the
 $SU(3)$ system including quarks. As is shown in Fig.1
for the case of quarks of
two light flavors at zero baryon density ($n_B=0$), the MP model
is consistent with lattice QCD data
providing a continuous phase transition of the cross-over   type
with a deconfinement temperature $T_{dec}=153$ MeV. 
 For a two-phase approach based on the bag model  a first-order 
deconfinement phase transition occurs with a sharp
jump in energy density $\varepsilon$ at $T_{dec}$ close to the 
value obtained from lattice QCD.

\begin{figure}[htb]
\begin{center}
\leavevmode
\epsfxsize=8.3cm
\epsfbox{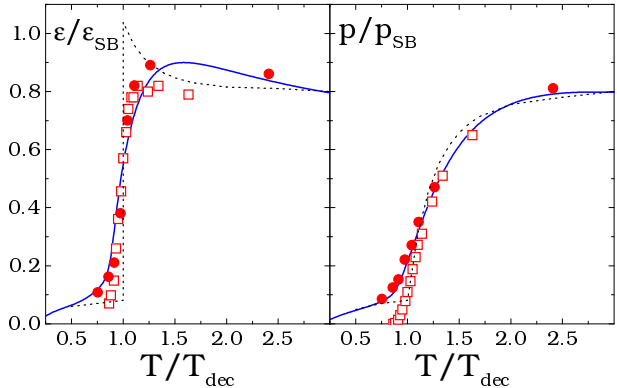}
\end{center}
\caption[C1]{ The reduced energy density $\varepsilon/\varepsilon_{SB}$ 
and pressure $p/p_{SB}$ (the $\varepsilon_{SB}$ and $p_{SB}$
are corresponding Ste\-phan-Boltz\-mann quantities) 
 of the $SU(3)$ system with two light flavors for $n_B=0$ calculated
within the MP (solid lines) and bag (dashed lines) models.
Circles and  squares are  lattice QCD data obtained within
the  Wilson~\cite{RS86} and Kogut--Susskind~\cite{berna} 
schemes, respectively. }
  \label{fig1}
\end{figure}

Though at a glimpse the temperature dependences of the energy 
density $\varepsilon$ and pressure $p$ for the different 
approaches presented in Fig.1
look quite similar, there are large differences revealed when
$p/\varepsilon$ is plotted versus $\varepsilon$ 
(cf. Fig.2, left panel).  The lattice QCD data 
differ at low $\varepsilon$, which is due to difficulties within
 the Kogut--Susskind scheme~\cite{berna} in treating the 
hadronic sector. A particular feature in the MP model is that, 
for $n_B=0$, the {\em softest point} of the EoS, 
defined  as a minimum of the function  $p(\varepsilon)/\varepsilon$
\cite{HS95}, is not very pronounced and 
located at comparatively  low values of the energy
density: $\varepsilon_{SP} \approx 0.45$ GeV/fm$^3$, which roughly
agrees with the lattice QCD value~\cite{RS86}.  This value of
$\varepsilon $ is close to the energy density inside the nucleon, and hence,
reaching this value indicates that we are dealing with  a single 
{\em big hadron} consisting of deconfined matter. In  contradistinction,  
the bag-model EoS exhibits a very pronounced softest point at large 
energy density $\varepsilon_{SP} \approx 1.5$ GeV/fm$^3$~\cite{HS95,R96}.

\begin{figure}[htb]
\begin{center}
\leavevmode
\epsfxsize=8.3cm
\epsfbox{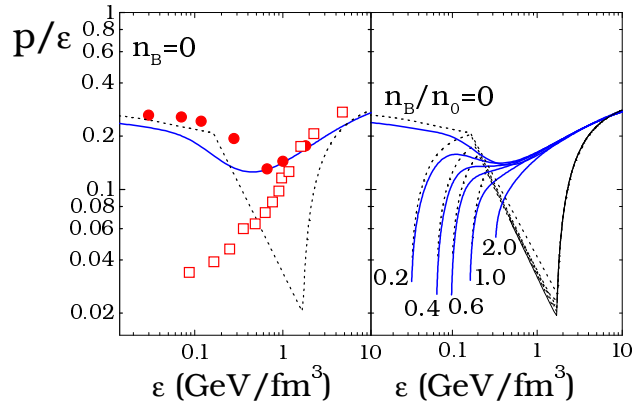}
\end{center}
\caption[C2]{The ($\varepsilon ,p/\varepsilon$)-representation of 
the EoS  
for the two-flavor $SU(3)$ system at various baryon densities 
$n_B$. Notation of data points and lines is the same as in Fig.1. }
\label{fig2}
\end{figure}

The MP model can be  extended to
baryon-rich  systems in a parameter-free way \cite{NST98,TNS98}. As
demonstrated in Fig.2 (right panel), the softest point for baryonic
matter is gradually washed out with increasing  baryon density
and vanishes for  $n_B \gsim 0.4 \ n_0$ ($n_0$ is
normal nuclear matter density). This behavior
differs drastically from  that of the two-phase bag-model EoS,  
where $\varepsilon_{SP}$ is only weakly dependent on 
$n_B$~\cite{HS95,R96}. It is of interest to note that
 the interacting hadron gas model has  no softest point at all and, 
in this respect, its thermodynamic behavior is close to that 
of the  MP model at high energy densities~\cite{TNS98}.

 These differences between the various models should manifest 
themselves in the  dynamics discussed below.

\section{Two-fluid hydrodynamic model}

In contrast to the one-fluid hydrodynamic model, where local 
instantaneous stopping of projectile and target matter is assumed,
a specific feature of the dynamical two-fluid description is
a finite stopping  power.  Experimental rapidity distributions 
in nucleus--nucleus  collisions support this specific feature
of the two-fluid model.
In accordance with \cite{MRS91}, the total baryonic current and
energy-momentum tensor are  written as
   \begin{eqnarray}
J^{\mu} &=& J^{\mu}_p + J^{\mu}_t~~,
   \label{eq7.1}
\\ 
T^{\mu\nu} &=& T^{\mu\nu}_p + T^{\mu\nu}_t~~,
   \label{eq7}
   \end{eqnarray}
where  the baryonic current  
$J^{\mu}_{\alpha}=n_{\alpha}u_{\alpha}^{\mu}$ 
and energy-momentum tensor $T^{\mu\nu}_{\alpha}$  of the fluid
$\alpha$ are initially associated with either target ($\alpha=t$) or 
projectile ($\alpha=p$) nucleons. Later on -- while heated up --  
 these fluids contain all hadronic and quark--gluon species, 
depending on the model used for describing the fluids. The twelve 
independent quantities (the baryon densities $n_{\alpha}$, 
4-velocities  $u_{\alpha}^{\mu}\;$ normalized as
  $u_{\alpha\mu}u_{\alpha}^{\mu}=1$, 
as well as temperatures and pressures of the fluids) are
obtained by solving the following set 
of equations of two-fluid hydrodynamics~\cite{MRS91}
   \begin{eqnarray}
   \label{eq8}
   \partial_{\mu} J_{\alpha}^{\mu} &=& 0~~, \\
   \partial_{\mu} T^{\mu\nu}_{\alpha} &=& F_{\alpha}^\nu~~,
   \label{eq8a}
   \end{eqnarray}
where the coupling term 
 \begin{eqnarray}
 F_{\alpha}^\nu=n^s_p n^s_t
\left < V_{rel} \int d\sigma_{NN\to NX}(s)\; (p - p_{\alpha})^\nu \right > 
 \label{eq9}
\end{eqnarray}
characterizes friction between the  counter-streaming fluids. 
Here,  $n^s_{\alpha}$ and $(p-p_{\alpha})$ denote respectively 
the scalar density 
of the fluid and  the 4-momentum transfer gained 
by a particle of the  fluid $\alpha$ after collision 
with a particle of the counter-streaming fluid. 
The cross sections $d\sigma_{NN\to NX}$ take into account all
elastic and inelastic interactions between the  constituents 
of different fluids at the invariant
collision energy $s^{1/2}$ with the local
relative velocity $V_{rel} =[s(s-4m_N^2)]^{1/2}/2m_N^2~.$ 
The average in (\ref{eq9}) is taken over  all particles 
in the two  fluids which are
assumed to be in local equilibrium intrinsically~\cite{MRS91}.
The set of Eqs.~(\ref{eq8}) and (\ref{eq8a}) is closed by an EoS, which
is naturally the same for both colliding fluids. 

The friction term $F^\nu_{\alpha}$ in Eq.~(\ref{eq8a}) originates  
from both elastic and inelastic $NN$ collisions. 
The latter give rise to  a direct
emission of mesons in addition to the thermal mesons in  the fluids. 
In the present version only for the pions the direct emission is 
included by the additional  equations
\begin{eqnarray}
\label{eq10}
\partial_\nu J_{\pi}^\nu &=& n^s_p n^s_t \left <
V_{rel} \int d\sigma_{NN\to \pi X} \right >~~,  \\
\partial_\mu  T^{\mu\nu}_{\pi} &=& n^s_p n^s_t \left < V_{rel}
\int d\sigma_{NN\to \pi X} \; p_{\pi}^\nu \right >~~, 
\label{eq10a}
\end{eqnarray}
where $p_{\pi}$ is the 4-momentum of an emitted direct pion. 
These equations together with (\ref{eq8a}) provide the  total 
energy--momentum conservation 
\begin{eqnarray} \partial_\mu (T^{\mu\nu}_{\pi} +
T^{\mu\nu}_{p} + T^{\mu\nu}_{t}) = 0~~. 
\label{eq11}
\end{eqnarray}
It is assumed~\cite{MRS91} that in the subsequent evolution
these direct  pions interact neither with the fluids nor with each
other. This is a reasonable  assumption at relativistic energies,
simulating a long formation time of these direct pions. 
At moderate energies, where the latter argument does not hold in general, 
the  number of direct pions is negligible compared to the number 
of thermal pions.

For the calculation of the friction force (\ref{eq9}), 
approximations of  $N$-$N$ cross-sections are used. 
It was found~\cite{Sat90} that a part of the friction term, which
is related to the transport cross-section, may be well 
parametrized  by  an effective deceleration length 
$\lambda_{\rm eff}$ with a constant value 
$\lambda_{\rm eff}\approx 5 $ fm. However, there are reasons 
to consider $\lambda_{\rm eff}$ as a phenomenological 
parameter, as it was pointed out in
\cite{MRS91}. First, the value of $\lambda_{\rm eff}$ is highly 
sensitive to the precise form of parameterization of the free 
cross-sections which, in addition, may be essentially 
modified by in-medium effects. Furthermore, the model neglects 
the interactions of direct pions both with each other
and with baryons, as well as  the direct emission of other 
mesons which are produced quite abundantly at  SPS energies. 
Due to all these effects   the stopping power at  SPS energies 
is somewhat underestimated~\cite{MRS91}. This shortcoming of
the model is cured by an appropriate choice of the 
$\lambda_{\rm eff}$ value as
$$ \lambda_{\rm eff} = a \; \exp(-b\sqrt{s}) $$
with  $a=6.6$ fm and $b=0.106$ GeV$^{-1}$  
 adjusted to  the rapidity distributions of nucleons and 
pions in central Au+Au  collisions at AGS and SPS energies. 

Following the original paper~\cite{MRS91}, it is assumed 
 that a fluid element decouples from the hydrodynamic regime, 
when its baryon density $n_B$ and densities in the eight
surrounding cells become smaller than a fixed value  $n_f$. A value
$n_f = 0.8 n_0$ was used for this local freeze-out density 
which corresponds to the actual density of the frozen-out  
 fluid element of about 0.5$n_0$ to 0.7$n_0$.

\section{Collective flow from heavy-ion collisions}

For central nucleus--nucleus collisions only the 
isotropic transverse expansion, or transverse radial flow, 
develops due to the azimuthal symmetry of a system. 
The presence of the reaction plane for non-central collisions
destroys  this symmetry  and gives rise to
 various patterns of collective motion generated by compressed and
excited nuclear matter created during the collision. For example,
the directed (or sideward) flow characterizes the deflection  
of emitted  hadrons away from the beam axis
within the reaction plane. In particular, one defines the 
differential directed flow by the mean in-plane
component $\left< p_x(y)\right>$ of the transverse momentum 
at a given rapidity $y$.  This deflection is believed to be 
quite sensitive to the {\em elasticity} or {\em softness} of the EoS.

The $\left< p_x(y)\right>$ distributions of  baryons are shown in Fig.3  
for Au+Au collisions
at $E_{lab}=10$ A$\cdot$GeV calculated for different EoS at an impact
parameter $b=3$ fm. In general,  the characteristic $S$-shape of the
distribution is reproduced, demonstrating a definite anti-correlation between
 nucleons bounced-off from the target and projectile regions. 
  One should keep in mind that the protons bound in
observed complex particles ({\em e.g.}, in deuterons) are not excluded 
in our calculations. Therefore, all hydrodynamic results 
should be compared to the triangle points in
Fig.3, where nucleons from complex particles do contribute. 
 The MP and interacting hadronic models\footnote{The interaction 
in the hadron model is
taken into account in the same manner as that in the hadronic sector of the MP
model~\cite{NST98,TNS98}.} give similar results, both getting into error bars
of these  triangle points, though
the flow in the MP model is slightly lower due to softening of EoS near 
the crossover phase transition. This softening is stronger for the bag-model  
EoS.  However, one should note  that  different versions of 

\begin{figure}[htb]
\begin{center}
\leavevmode
\epsfxsize=8.3cm
\epsfbox{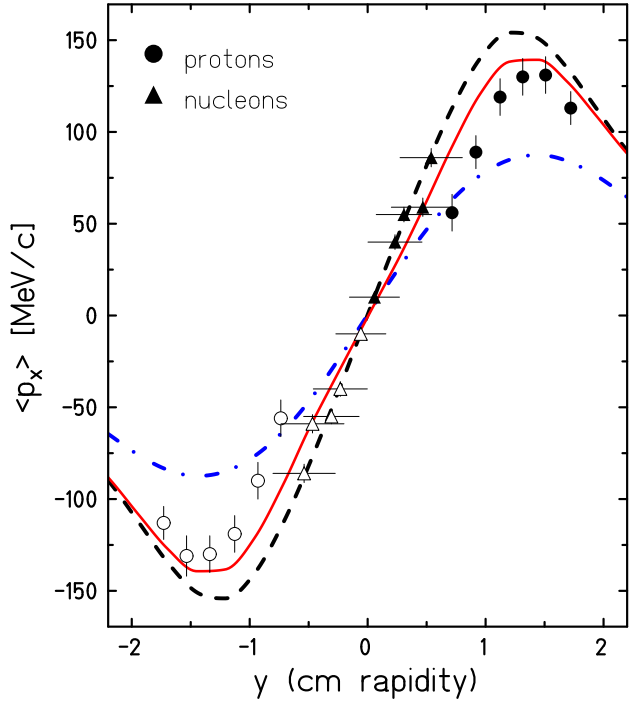}
\end{center}
\caption[C3]{ Differential directed flow of nucleons in the 
reaction plane as a function of ra\-pi\-di\-ty in semi-central (the trigger
transverse energy $E_T =$ ($200-230$) GeV) Au + Au collisions 
at the energy $10$ A$\cdot$GeV.  Three curves are calculated 
within  relativistic two-fluid hydrodynamics for an impact parameter
$b=3$ fm and different EoS: for the MP model (solid line),
interacting hadron gas model (dashed) and two-phase bag model (dot-dashed).
Circles are experimental points for  identified protons, triangles
correspond to a nucleon flow estimate based on the measurement of $E_T$
and the number of charged particles $N_C$~\cite{exp10}.  Experimental points
marked by full symbols are measured directly, open ones are obtained by
reflecting at the mid-rapidity point. }
\label{fig3}
\end{figure}
\noindent  transport codes, which do not  
  account  for a phase transition, also give a reasonable 
description of 
$\left< p_x(y)\right>$ 
({\em e.g.}, see the comparison with RQMD  results 
in~\cite{exp10}). Therefore, convincing evidence 
on a possible phase transition, based solely on the data 
at a single bombarding energy, is hardly possible.

The rapidity dependence of the mean in-plane transverse momentum can be
quantified in terms of the derivative at mid-rapidity
\begin{equation}
\label{eq12}
F_y = \left. \frac{d \; \left<p_x(y)\right>}{dy} \right|_{y=y_{cm}}~~, 
\end{equation}
which is quite suitable for analyzing the flow excitation function. The
slope parameter $F_y$ calculated with the MP-model EoS is presented 
in Fig.4 (upper panel) together with available
experimental points covering the whole range of incident
energies. The results describe correctly the decrease of  $F_y$
with increasing energy and shows essentially no dependence  on
 impact parameter for  semi-central collisions.  
It was shown experimentally~\cite{FOPI} that the
directed flow is larger for heavier fragments. As  mentioned before, 
hydrodynamic calculations deal with {\em primordial} nucleons, and hence 
they describe the mean value of $F_y$ for free nucleons and 
nucleons bound in deuterons and heavier fragments. Therefore, 
 our hydrodynamic results lie between the
experimental points for identified protons (open circles in Fig.4) 
and the data~\cite{FOPI} (full circles in Fig.4)  for intermediate 
mass fragments\footnote{For complex particles the value $F_y$ was
deduced from $p_x$ per baryon. The FOPI data for intermediate
mass fragments are often scaled  by factor 0.7 to make absolute values
comparable to those for $p,d,\alpha$~\cite{FOPI}. We do not use this 
scaling factor. Note that everywhere we deal  with $F_y$ defined 
by Eq.(\ref{eq12}), which differs from the frequently used flow 
parameter $F=F_y \ y_{cm}$. The extra beam-rapidity factor $y_{cm}$
obscures a relative role of dynamical effects in the energy dependence 
of the directed-flow slope parameter. }.  
This effect  is particularly strong for energies below 
$E_{lab}\approx  1$~A$\cdot$GeV, where the baryonic flow is largest 
(and  heavy fragments in the mid-rapidity range are abundant).

To our knowledge no other hydrodynamic calculations of $F_y (E_{lab})$  
have been reported. Therefore, we compare our results 
with transport calculations  
in the lower panel of Fig.4. The ARC and ART are cascade models, 
while the RQMD takes also into account mean-field effects.
Though all these models agree with experimental data at $E_{lab} \approx 10$
A$\cdot$GeV (considered as a reference point), values of $F_y$ at
lower energies are clearly underestimated, as is evident from comparison 
with preliminary results of the E895 Collaboration~\cite{E895} (see
empty squares in Fig.4). Recently, a good description of
 experimental points (including the E895 data) was
reported within a relativistic BUU  (RBUU) model~\cite{cas99}. 
  The good agreement with experiment was achieved by
a special fine tuning of the mean fields involved in the particle 
propagation.
  
\begin{figure}[t]
\begin{center}
\leavevmode
\epsfxsize=8.3cm
\epsfbox{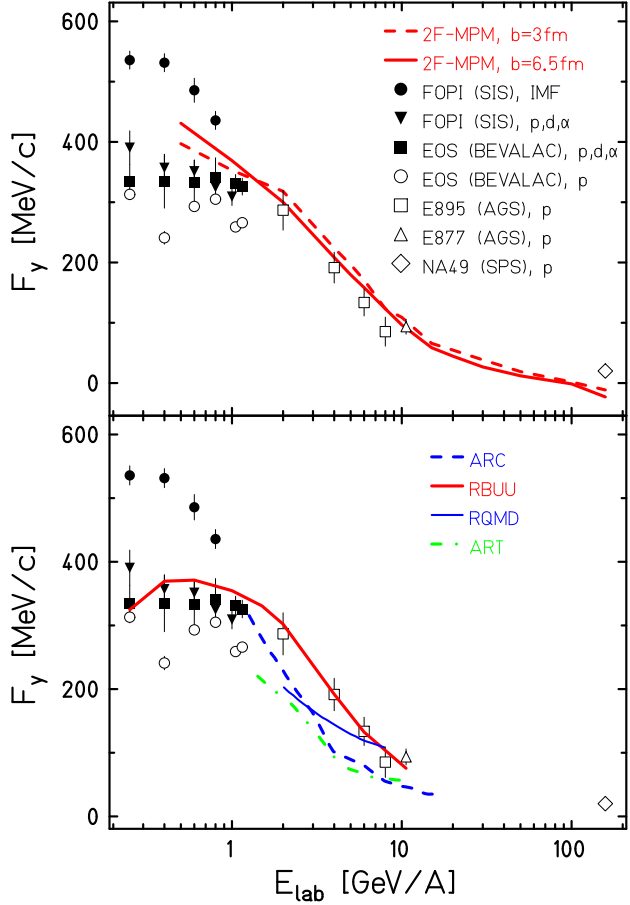}
\end{center}
\caption[C4]{Excitation function of the slope parameter $F_y$ for baryons
 from Au + Au collisions within two-fluid hydrodynamics for the MP
EoS (upper panel) and  within different transport simulations (lower panel).
Open symbols are experimental points for identified protons
(see data collection in~\cite{Dan99,FOPI,Aj98}), filled circles, triangles
and squares correspond to the flow parameter measured for intermediate mass
fragments~\cite{FOPI} and for light particles 
$p, d, \alpha$~\cite{Her96,EOS}. The
results of transport calculations for three different codes are given by the 
thin solid  (RQMD), dashed (ARC) and dot-dashed (ART) lines (cited according
to~\cite{Aj98}) . The solid line (RBUU) is taken from~\cite{cas99} }
\label{fig4}
\end{figure}

It is of interest to mention that the calculated value of 
$F_y$ for the baryon flow 
becomes negative (antiflow) for beam energies  $\gsim 100$ A$\cdot$GeV, 
while the experiment~\cite{NA49} gives a small but positive value 
even at 158 A$\cdot$GeV.  The reason of this antiflow is a wiggle 
in the $\left<p_x(y)\right>$ 
distribution arising in hydrodynamic results within 
a narrow mid-rapidity interval 
$|\delta y| \lsim 1$ due to a peculiar interplay between the 
transverse radial and 
directed flows. The possibility of such an effect was noticed  
 in~\cite{voloshin} some time ago and later also observed in the 
UrQMD transport calculations~\cite{URQMD}. 
However, actual measurements have been taken at
larger rapidities and then extrapolated into this unmeasured
region~\cite{NA49}. Therefore, more accurate data in the 
mid-rapidity region are necessary to clarify this behavior. 

The directed flow can be characterized by another quan\-tity which is 
less sensitive to possible rapidity fluctuations of the 
in-plane momentum. Such a quantity is the average directed flow 
which is defined by
\begin{equation}
\left< P_x\right> = \frac{\displaystyle\int dp_xdp_ydy \ p_x \ 
\left(  E{\displaystyle 
\frac{d^3N}{dp^3}}\right)}{\displaystyle\int  dp_xdp_ydy \ \left(
    E{\displaystyle 
 \frac{d^3N}{dp^3}}\right) }~~,
\label{eq13}
\end{equation}
where the integration in the c.m.system runs over the rapidity region 
$\displaystyle [0,y_{cm}]$. The calculated excitation functions 
for the average directed flow of baryons within different 
models are shown in Fig.5. Conventional (one-fluid) hydrodynamics 
 for pure hadronic matter~\cite{R96} results 
\begin{figure}[b]
\begin{center}
\leavevmode
\epsfxsize=8.3cm
\epsfbox{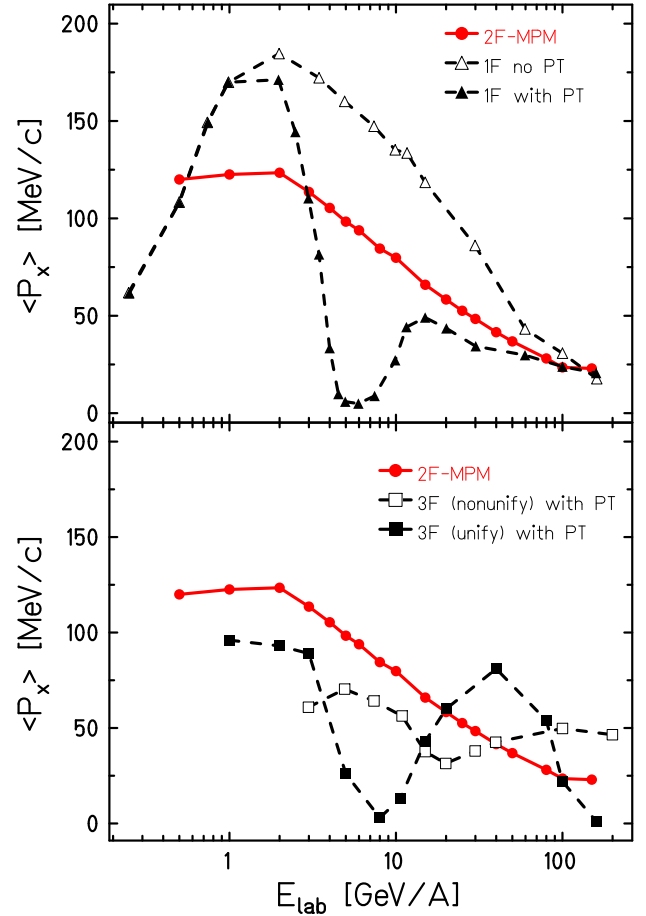}
\end{center}
\caption[C5]{The excitation function of the average directed flow for baryons
from Au + Au collisions. Two-fluid hydrodynamics with the MP
EoS at the impact parameter 3 fm is compared with the corresponding results
of  one-fluid~\cite{R96} (upper panel) and  three-fluid 
(lower panel)~\cite{3DF} hydrodynamics with the 
bag-model EoS.  One-fluid calculations both with 
and without the phase transition (PT) are displayed. }
\label{fig5}
\end{figure}
\noindent in a  very large directed flow due to the inherent 
instantaneous stopping of the colliding matter. This 
instantaneous stopping is unrealistic at high beam energies. If the
 deconfinement phase transition, based on the bag-model EoS~\cite{R96}, 
is included into this model, the excitation function  
of $\left< P_x\right>$ exhibits a
deep minimum near $E_{lab}\approx 6$  A$\cdot$GeV, which manifests the
softest-point effect of the bag-model EoS depictured in the right panel 
of Fig.2.

The result of  two-fluid hydrodynamics with the MP EoS noticeably differs
from the one-fluid calculations. After a maximum around 1 A$\cdot$GeV, 
the average directed flow  decreases slowly and smoothly. This
difference is caused by two reasons. First, as  follows from Fig.2, the
softest point of the MP EoS is washed out for $n_B \gsim 0.4$. The second
reason is dynamical: the finite stopping power and direct 
pion emission change the evolution pattern. The latter point is confirmed by
comparison to three-fluid calculations with the bag EoS~\cite{3DF}
plotted in the lower panel of Fig.5. 
 The third pionic fluid in this model
is assumed to interact only with itself neglecting the interaction with
baryonic fluids. Therefore, with regard to 
the baryonic component, this three-fluid hydrodynamics~\cite{3DF,3DF97} is
completely equivalent to our two-fluid model and the main
difference is due to the different EoS. As seen in Fig.5, the minimum of the
directed flow excitation function, predicted by the one-fluid hydrodynamics 
with the bag  EoS, survives in the three-fluid (nonunified) 
regime but its value decreases and its position shifts to 
higher energies. If one  applies the {\em unification 
procedure} of ~\cite{3DF}, which favors fusion of two fluids 
into a single one, and thus making stopping larger, three-fluid hydrodynamics 
practically reproduces  the one-fluid  result and predicts 
in addition a bump at $E_{lab}\approx 40$  A$\cdot$GeV.

 \section{Conclusions}

We have studied relativistic nuclear collisions within  3D two-fluid
hydrodynamics combined with different EoS, including that of the statistical
mixed-phase model of the deconfinement phase transition, developed 
in~\cite{NST98,TNS98}. It has been shown that the directed flow excitation
functions $F_y$ and $\left< P_x\right>$ for baryons  
 are sensitive to the EoS, but this sensitivity 
is significantly masked  by nonequilibrium dynamics of nuclear collisions.
Nevertheless, the  results indicate that the widely used
two-phase EoS, based on the  bag model~\cite{HS95,R96} and 
giving rise to a first-order phase transition, seems to be 
inappropriate. The  neglect of
interactions near the deconfinement temperature results in an
unrealistically  strong softest-point effect within this two-phase EoS.
In fact, its prediction of a minimum in $\left< P_x\right>(E_{lab})$
near $E_{lab}\approx 6$ A$\cdot$GeV has not been confirmed 
experimentally~\cite{E895}. However, accurate experimental investigations of
the differential directed flow   and flow excitation functions in
the energy region between AGS and SPS are still highly demanded  not
only in searching for a shifted minimum of 
$\left< P_x\right>(E_{lab})$, but also in
clarifying the physics of a possible  negative slope (antiflow) of the
baryonic directed flow $F_y$.
This antiflow is particularly sensitive to the EoS. 
While  for the EoS in the MP model the antiflow is predicted 
 at incident energies only above 100 A$\cdot$GeV, it occurs already at 8 
A$\cdot$GeV, when the bag EoS is used~\cite{3DF}. 

In this respect  a dramatic phenomenon of
the {\em cracked nut}  proposed recently as a hydrodynamic 
signature of the QCD phase transition  at  RHIC and LHC 
energies~\cite{TS99,KSH99} looks questionable. The authors   argue 
that the softest point in the EoS may lead to the development 
of two shells at the edge of the almond-like initial
fireball, which are then cracked by internal pressure and 
separated, resulting in a specific flow pattern.
However, this speculation was based on the bag-model EoS.  
The application of the QCD-consistent EoS of the MP
model to this problem would be  interesting.

The directed flow is  the first coefficient in the Fourier
decomposition of the azimuthal momentum distribution of
particles~\cite{PV98}.   The second coefficient, the elliptic flow, is
expected to be  more sensitive to the EoS and some hints of the phase
transition have been indicated  by an analysis of the measured
excitation function for the elliptic flow (see review
articles~\cite{BMS98,Dan99}). The study of the elliptic flow within
 two-fluid hydrodynamics for the mixed-phase model EoS is
in progress. \\

 \acknowledgements

We are grateful to V.~Russkikh for making his hydrodynamic code available
to us. We thank P.~Danielewicz, B.~Friman and E.~Kolomeitsev for useful 
discussions.  Yu.B.I., E.G.N. and
V.D.T.  gratefully acknowledge the hospitality at the Theory Group of GSI,
where  this work has been done.
This work was supported in part by DFG (project 436 RUS 113/558/0) and RFBR 
(grant 00-02-04012). 
Yu.B.I. was partially supported by RFBR grant 00-15-96590.

 \end{document}